\documentstyle[graphicx, 12pt]{article}
\begin{document}

\small
\hoffset=-1truecm
\voffset=-2truecm
\title{\bf The circular loop equation of a cosmic string with time-varying tension }
\author{Hongbo Cheng\footnote {E-mail address: hbcheng@public4.sta.net.cn} \hspace {1cm} Yunqi Liu\\
Department of Physics, East China University of Science and
Technology,\\ Shanghai 200237, China}

\date{}
\maketitle

\begin{abstract}
The equation of circular loops of cosmic string with
time-dependent tension is studied in the Minkowski spacetime and
Robertson-Walker universe. We find that, in the case where the
tension depends on some power of the cosmic time, cosmic string
loops with time-varying tension should not collapse to form a
black hole if the power is lower than a critical value.
\end{abstract}
\vspace{10cm}
\hspace{1cm}
PACS number(s): 98.80.Cq

\newpage
More than twenty years ago, as a kind of topological defects,
cosmic strings including their formation, evolution and
observational effects, started to attract much attention [1-3]. As
linear defects at a symmetry breaking phase transition in the
early universe, cosmic strings can be produced at the end of
inflation. They have several potentially important astrophysical
features. In particular, one is providing an explanation for the
origin of the primordial density perturbations leading to the
existence of galaxies and clusters. Another is their spacetime
metric with a deficit angle which could give rise to some
observational results. However, the improved observational data
like CMB anisotropy observations and the WMAP experiments rule out
cosmic strings as seeds of large scale structure formation in the
universe because of limits on their tension ($G\mu\leq10^{-6}$)
[4-6].

Recently there has been a resurgence of interest in astrophysical
results of cosmic strings for both theoretical and observational
reasons although cosmic strings can not play an important role in
the formation of large scale structure in our universe.
Topological defects, including cosmic strings, are inevitably
formed at the end of inflation. They can also generate at the end
of brane inflation [7, 8], which provides us with a potential
window on M theory [9-11]. Furthermore cosmic strings still have
strong influence on various astrophysics [1-3] such as
gravitational lensing effects [12, 13], gravitational wave
background [14, 15], early reionization [16, 17] and so on. A
gravitational lens called CSL-1 invoking two imagines of
comparable magnitude of the same giant elliptical galaxy were
discovered. It is interesting that many similar objects were found
in the vicinity [18, 19]. The cosmic string can become an
important and powerful explanation for these experimental
findings.

It is necessary to explore the evolution of cosmic string loops
extensively and deeply. Once cosmic string formed at any epoch in
the history of the universe, they are not static and would envolve
under their own force of tension continuously instead. They can
collide and intersect to undergo reconnections although the
strings stretch under the influence of the Hubble expansion or the
environment and the strings lose energy to gravitational radiation
when they oscillate. The reconnections of long strings and large
loops will produce small loops copiously. In general the methods
of existence of cosmic strings are string network consisting of
long strings and closed string loops. As a complicated time
dependent gravitational source, the cosmic string loops oscillate
with time  rather randomly. On the experimental side, Schild et al
observed and analysed the anomalous brightness fluctuation in the
multiple-image lens system Q0957+561A, B which has been
investigated for 25 years [20, 21]. They think that system
consists of two quasar images separated by approximately $6''$.
The phenomena are known to be images of the same quasar not only
because of the spectroscopic match, but also because the images
fluctuate in brightness, and the time delay between fluctuations
is always the same. The effect may be due to lensing by an
oscillating loop of cosmic string between us and the lensing
system, because loops of cosmic strings supply a quantitative
explanations of such synchronous variations in two images. The
cosmic string loops can also generate more gravitational waves and
distinct signatures [14, 15, 22]. In a word once the cosmic
strings appeared, certainly they evolve to produce the cosmic
string loops supported by the possible examples including those
above. The evolution and fate of cosmic string loops attracted
more attention. More efforts have been paid for the research in
various backgrounds. In some cases like Minkowski spacetime and
Robertson-Walker universe, the loops will collapse to form black
holes or become a long cosmic string instead of remaining
oscillating loops [23, 24]. In de Sitter universe, only loops with
larger initial radii can survive [23, 25]. In the Kerr-de Sitter
environment, around rotating gravitational sources with positive
cosmological constant, a lot of cosmic string loops, including
smaller ones, can evolve to survive when the gravitational source
rotates faster [26]. According to the examples mentioned above, it
is clear that most of cosmic string loops will become black holes
unless the loops live around the rotating gravitational source
with greater angular momentum in the de Sitter spacetime. It
should be pointed out that all results mentioned above are
obtained when the tension of cosmic string is chosen to be
constant.

It is interesting and significant to investigate the evolution of
cosmic string loops with time-dependent tension. So far in nearly
all researches, that the tensions of cosmic string are constant is
just an assumption. In cosmological situations the cosmic strings
with time-varying tension can often appear. M. Yamaguchi put
forward the important issue [27]. The tensions of cosmic strings
can depend on the cosmic time. For example, a potential containing
one complex scalar field $\phi$ and one real scalar field $\chi$
can be written as
$V(\phi,\chi)=\frac{\lambda}{4}(|\phi|^{2}-\chi^{2})^{2}+\frac{1}{2}m_{\chi}^{2}\chi^{2}$.
The backreaction to the oscillation of field $\chi$ is negligible
as the coupling constant $\lambda$ is sufficiently small. The
tension of string $\mu$ is associated with the root mean square of
the expection value of field $\chi$ and can be denoted as
$\mu\propto a^{-3}$ where $a$ is the scale factor which is
proportional to $t^{\frac{1}{2}}$ in the radiation-dominated era
and $t^{\frac{2}{3}}$ in the matter-dominated era [27]. Some works
on this topic are performed and the interesting and important
conclusions are drawn [27, 28]. In the case where the tension
depends on the power of the cosmic time assumed as $\mu\propto
t^{q}$, such cosmic strings go into the scaling solution when
$q<1$ in the radiation domination and $q<\frac{2}{3}$ in the
matter domination. They also pointed out that the CMB and matter
power spectra induced by cosmic strings with time-dependent
tension can be different significantly from those generated by the
conventional cosmic strings with constant tension, which encourage
us to consider that a lot of related topics need to be
investigated. Until now little contribution is made to scrutinize
the evolution of cosmic string loops which possess time-dependent
tension.

The purpose of this paper is to obtain the equation of circular
loops of cosmic string with time-varying tension in the Mionkowski
spacetime and Robertson-Walker universe. We wonder the time
dependence of the tension on the evolution and fate of the cosmic
string loops. First of all we derive the equations of circular
loops of cosmic string in the expanding universe by means of the
Nambu-Goto action with an additional factor for the time-dependent
tension. We solve the equations numerically to study the evolution
of loops and the time dependence of tension on the fate of cosmic
string loops. Finally the conclusions and discussions are
emphasized.

We start to consider the evolution of cosmic string loops whose
tensions are functions of cosmic time in the expanding universe.
The Robertson-Walker metric is written as,

\begin{equation}
ds^{2}=dt^{2}-R^{2}(t)(dr^{2}+r^{2}d\theta^{2}+r^{2}\sin^{2}\theta
d\varphi^{2})
\end{equation}

\noindent where the scale factor is,

\begin{equation}
R(t)=R_{0}t^{\beta}
\end{equation}

\noindent here we choose $\beta=\frac{1}{2}$ for
radiation-dominated era and $\beta=\frac{2}{3}$ for
matter-dominated era. The Nambu-Goto action for a cosmic string
with time-dependent tension is given by,

\begin{equation}
S=-\int d^{2}\sigma\mu(t)[(\frac{\partial
x}{\partial\sigma^{0}}\cdot\frac{\partial
x}{\partial\sigma^{1}})^{2}-(\frac{\partial
x}{\partial\sigma^{0}})^{2}(\frac{\partial
x}{\partial\sigma^{1}})^{2}]^{\frac{1}{2}}
\end{equation}

\noindent where $\mu(t)$ is the string tension and the function of
cosmic time. $\sigma^{a}=(t,\varphi)$ ($a=0,1$) are timelike and
spacelike string coordinates respectively. $x^{\mu}(t,\varphi)$
($\mu,\nu=0,1,2,3$) are the coordinates of the string world sheet
in the spacetime.

For simplicity let us assume that the string sheet we study lies
in the hypersurface $\theta=\frac{\pi}{2}$, then the spacetime
coordinates of the world-sheet parametrized by $\sigma^{0}=t$,
$\sigma^{1}=\varphi$ can be chosen as
$x=(t,r(t,\varphi),\frac{\pi}{2},\varphi)$.

In the case of planar circular loops, we have $r=r(t)$. According
to the metric (1) and the spacetime coordinates mentioned above,
the Nambu-Goto action with an additional factor for the
time-dependent tension of cosmic string denoted as (3) is reduced
to,

\begin{equation}
S=-\int dtd\varphi\mu(t)r(R-R^{2}\dot{r}^{2})^{\frac{1}{2}}
\end{equation}

\noindent which leads to the following equation of motion for
loops,

\begin{equation}
r\ddot{r}+\frac{d\ln\mu}{dt}r\dot{r}(1-R\dot{r}^{2})+\frac{1}{R}-\dot{r}^{2}
+\frac{3}{2}\frac{\dot{R}}{R}r\dot{r}-\dot{R}r\dot{r}^{3}=0
\end{equation}

\noindent In order to discuss the equation (5) carefully to
explore the evolution of this kind of cosmic string loops, we
represent the time-varying tension for simplicity as follow,

\begin{equation}
\mu(t)=\mu_{0}t^{q}
\end{equation}

\noindent then the equation of motion (5) becomes,

\begin{equation}
r\ddot{r}+\frac{q}{t}r\dot{r}(1-R\dot{r}^{2})+\frac{1}{R}-\dot{r}^{2}
+\frac{3}{2}\frac{\dot{R}}{R}r\dot{r}-\dot{R}r\dot{r}^{3}=0
\end{equation}

In the case of Minkowski spacetime, i.e. $R(t)=constant$, equation
(7) is reduced to,

\begin{equation}
r\ddot{r}+\frac{q}{t}r\dot{r}(1-R\dot{r}^{2})-\dot{r}^{2}+\frac{1}{R}=0
\end{equation}

\noindent If the tension of cosmic string keeps constant as $q=0$,
equation (8) has been solved by Vilenkin [2]. The analytical
solution $r(t)=r_{0}\cos\frac{t-t_{0}}{R_{0}}$ with
$r(t_{0})=r_{0}$ and $\dot{r}(t_{0})=0$ shows the collapsing of
all closed string loops. If the cosmic strings possess the
time-dependent tension changing as the power $q$ of time like (6),
equation (8) can be solved numerically. Having performed the very
burden and difficult calculation, we find that there must exist a
critical value $q_{f}=-0.131$. When $q<q_{f}$, all of cosmic
string loops will expand to evolve or contrarily will collapse to
form black holes when $q>q_{f}$ under $\dot{r}(t_{0})=0$ no matter
how large the initial values of loop radius $r(t_{0})=r_{0}$ is
equal to. We also discover that the cosmic string loops in the
case of time-dependent tension with $q>q_{f}$ contract faster than
ones in the case of constant tension and $q$ is not chosen to be a
negative integer only when $\dot{r}(t_{0})=0$. The evolution of
radii of circular loops with time-varying tension in the Minkowski
pacetime is depicted in Figure 1. Therefore we have now shown an
argument as follow. In the Minkowski spacetime, if the expression
of tension depending on some power of the cosmic time satisfies
the conditions such as $q<q_{f}$ for the case of $\mu(t)\propto
t^{q}$, the cosmic string loops expand to evolve instead of
becoming black holes.

Next, we shall discuss the Robertson-Walker universe for which the
scale factor $R(t)$ denoted as (2). By using (2) and (7), we have,

\begin{equation}
r\ddot{r}+\frac{q}{t}r\dot{r}(1-R_{0}t^{\beta}\dot{r}^{2})-R_{0}\beta
t^{\beta-1}r\dot{r}^{3}-\dot{r}^{2}+\frac{3}{2}\frac{\beta}{t}r\dot{r}+\frac{1}{R_{0}t^{\beta}}=0
\end{equation}

\noindent For the case of constant tension equivalent to $q=0$,
equation (9) can be reduced to,

\begin{equation}
r\ddot{r}-R_{0}\beta t^{\beta-1}r\dot{r}^{3}-\dot{r}^{2}
+\frac{3}{2}\frac{\beta}{t}r\dot{r}+\frac{1}{R_{0}t^{\beta}}=0
\end{equation}

\noindent This equation has been solved to show that all cosmic
string loops will collapse to become black holes at last in the
expanding universe although they will expand a little at first [2,
24]. For the loops of cosmic string whose tension changes as the
power of time, equation (9) can be solved numerically by means of
burden and surprisingly difficult calculation. We find that there
also exist a special value for every era, $q_{r}=-0.078$ for
radiation-dominated era and $q_{m}=-0.062$ for matter-dominated
era. In each era when $q<q_{r}$ or $q<q_{m}$ respectively, all
cosmic string loops will expand to evolve under $\dot{r}(t_{0})=0$
with any values of $r(t_{0})=r_{0}$ the initial values of loop
radius. Contrarily all loops will become black holes at last while
they also contract faster than ones with constant tension. In the
Robertson-Walker background the evolutions of radii of circular
loops with changeable tension in the radiation-dominated era and
matter-dominated era are depicted in Figure 2 and Figure 3
respectively. The power $q$ is chosen to be a negative value
excluding the negative integer only when $\dot{r}(t_{0})=0$. The
results in the Robertson-Walker universe are similar to those in
the Minkowski spacetime. There are only revisions of the critical
values for the power $q$ in comparison with ones in the Minkowski
spacetime. It should be stressed that $q_{r}>q_{f}$ and
$q_{m}>q_{f}$ appear because the expansion of universe leading the
loops enlarge. Therefore, we have now shown an argument as follow.
In the Robertson-Walker universe, when the expression of tension
depending on some power of the cosmic time satisfies the
conditions such as $q<q_{r}$ or $q<q_{m}$ for the case of
$\mu(t)\propto t^{q}$, the cosmic string loops expand to evolve
instead of becoming black holes. It was found that the cosmic
strings whose tension depends on some power of the cosmic time as
mentioned above relax into the scaling solution only when $q<1$ in
the radiation domination and $q<\frac{2}{3}$ in the matter
domination [27, 28]. Having compared our findings with the
conclusions in [27, 28], we also find that $q_{r}<1$ and
$q_{m}<\frac{2}{3}$, which means that such strings with $q<q_{r}$
or $q<q_{m}$ will form the expanding loops and can also go into
the scaling solution. Our results are consistent with those in
[27, 28].

According to the observational results, there could exist a lot of
cosmic string loops in our universe [20, 21]. Here we report the
evolution of planar circular loops of cosmic string with
time-varying tension in the Robertson-Walker universe,
investigating the fate of loops again. But in the case of cosmic
string with constant tension, more efforts were paid to show that
all loops will collapse to form black holes in the expanding
universe and most loops also become black holes in the de Sitter
spacetime except the larger ones, which means that only fewer
loops can survive. In the case of cosmic string with changeable
tension we let the tension is proportional to the power of time
like equation (6) for simplicity and without losing generality. We
discover that the loops with conditions $q<-0.078$ or $q<-0.062$
during radiation-dominated or matter-dominated era respectively
will expand to evolve instead of becoming black holes. When the
power is negative, the tension will be smaller and smaller and its
influence leading the loops contract become weaker and weaker. It
is clear that the fate of loops can change significantly if the
expression of tension of cosmic string obey the conditions which
is certainly different for different expressions. Our findings
indicate that there may exist a considerable number of loops of
cosmic string but their tension is a time function obeying the
specific conditions. We can keep observing the effect of cosmic
string loops.

The main results of this paper is equation (5), the circular loop
equation for a cosmic string evolving in the hypersurface with
$\theta=\frac{\pi}{2}$ in the Robertson-Walker universe. A
remarkable property about this equation is that a loop may never
contract to one with a Schwarzschild radius, if the expression for
tension of cosmic string satisfies the necessary conditions while
these cosmic strings must go into the scaling solution. Therefore
a lot of cosmic string loops can evolve to survive in our
universe. Our research is a starting point for the case of cosmic
string loops with changeable tension although we let the function
of tension is some power of cosmic time here. The evolution of
cosmic string loops possessing the changeable tension in other
spacetime requires further research.

\vspace{1cm}
\noindent \textbf{Acknowledge}

This work is supported by NSFC No. 10333020 and the Shanghai
Municipal Science and Technology Commission No. 04dz05905.

\newpage
\begin{figure}
\setlength{\belowcaptionskip}{10pt} \centering
  \includegraphics[width=15cm]{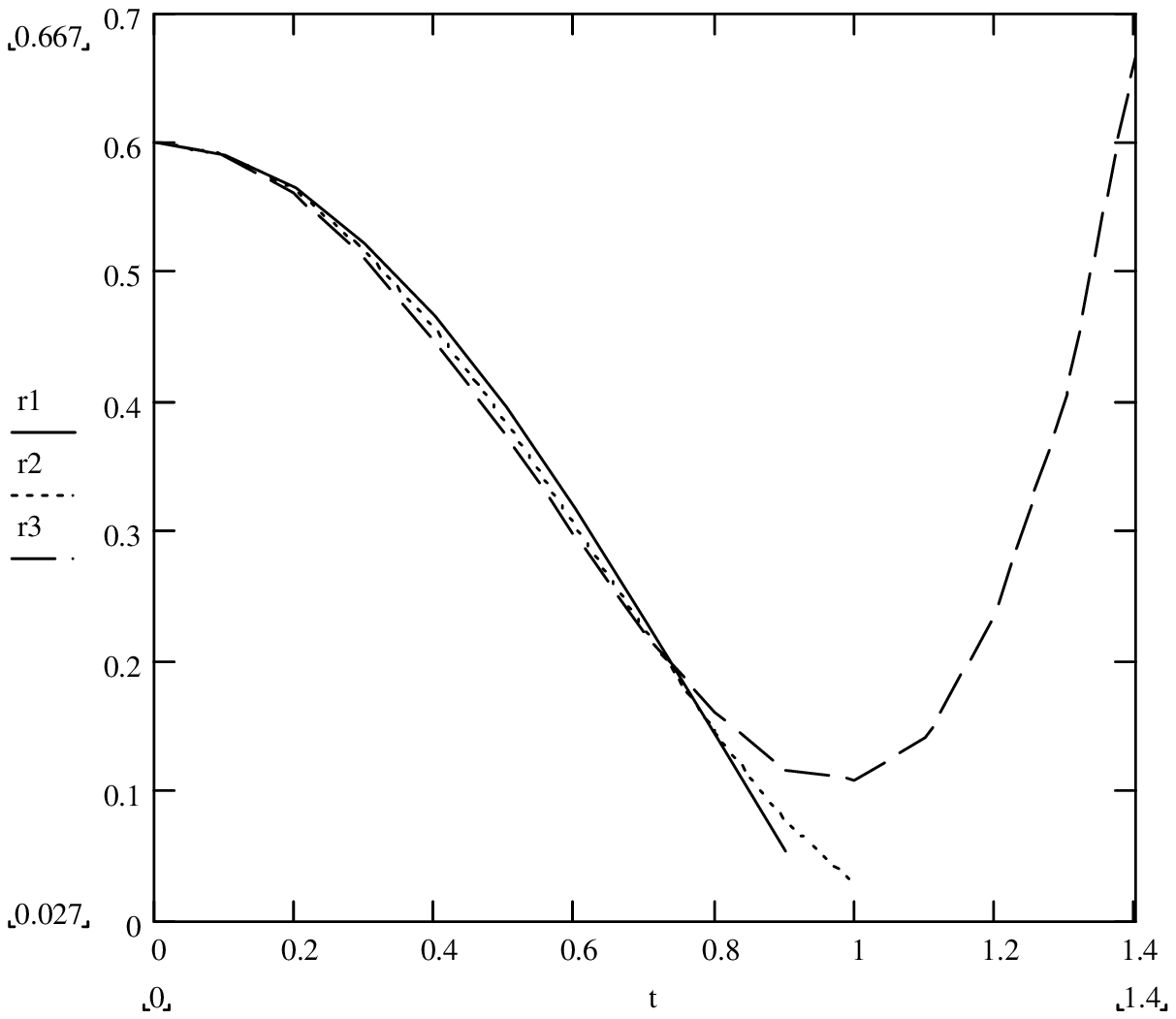}
  \caption{The solid, dot, dashed curves of $r(t)$ the radii of circular loops as functions of cosmic time with
  $q=-0.05, -0.13, -0.2$ respectively and initial value $r(t_{0})=0.6$ and $\dot{r}(t_{0})=0$ in the Minkowski spacetime.}
\end{figure}

\newpage
\begin{figure}
\setlength{\belowcaptionskip}{10pt} \centering
  \includegraphics[width=15cm]{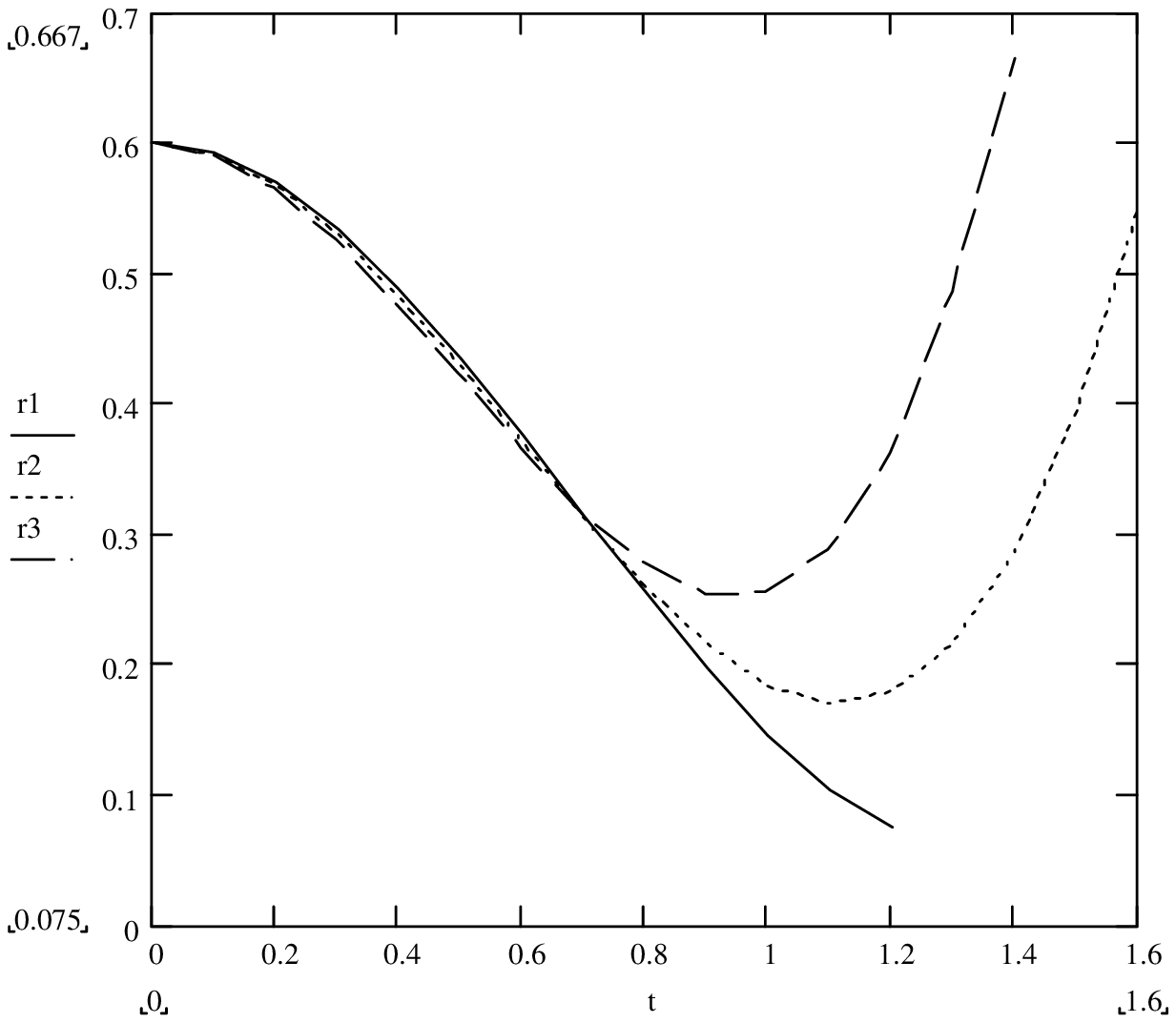}
  \caption{The solid, dot, dashed curves of $r(t)$ the radii of circular loops as functions of cosmic time with
  $q=-0.02, -0.08, -0.15$ respectively and initial value $r(t_{0})=0.6$ and $\dot{r}(t_{0})=0$ in the radiation-dominated era.}
\end{figure}

\newpage
\begin{figure}
\setlength{\belowcaptionskip}{10pt} \centering
  \includegraphics[width=15cm]{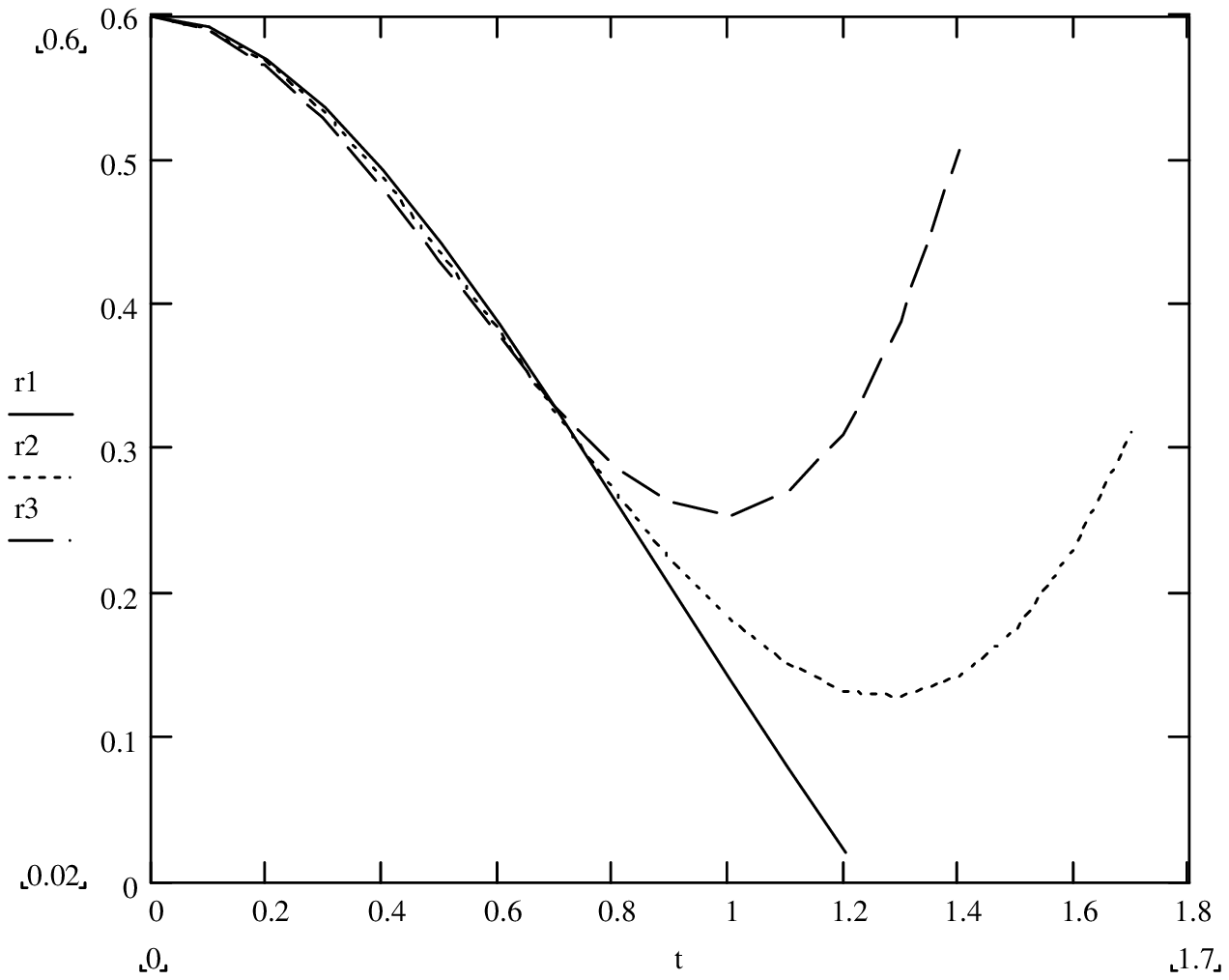}
  \caption{The solid, dot, dashed curves of $r(t)$ the radii of circular loops as functions of cosmic time with
  $q=-0.01, -0.08, -0.15$ respectively and initial value $r(t_{0})=0.6$ and $\dot{r}(t_{0})=0$ in the matter-dominated era.}
\end{figure}

\end{document}